\documentclass[runningheads]{llncs}

\usepackage{graphicx}
\usepackage{color}

\usepackage{listings}
\usepackage{lstautogobble}  

\usepackage{amssymb}
\usepackage{amsmath}

\usepackage{xspace}
\usepackage{hyperref}
\usepackage{multirow}
\usepackage{makecell}
\usepackage[caption=false]{subfig}
\usepackage{booktabs}

\usepackage[mode=buildnew]{standalone}
\usepackage{tikz}
\usetikzlibrary{positioning}
\usetikzlibrary{calc}
\usetikzlibrary{fit}
\usetikzlibrary{shapes.arrows}
\usetikzlibrary{backgrounds}

\usepackage{soulutf8}

\usepackage{tabularx}
\usepackage{arydshln}
\usepackage{pdflscape}
\usepackage{adjustbox}
\usepackage{float}

\usepackage[T1]{fontenc}

\newcommand{\TODO}[1]{
  \if\relax\detokenize{#1}\relax
    \textcolor{red}{TODO}
  \else
    \textcolor{red}{ {#1}}
  \fi
}

\newcommand\addressID{\ensuremath{a}\xspace}
\newcommand\addressIDn[1]{\ensuremath{\addressID_#1}\xspace}
\newcommand\addressPairs[1]{\ensuremath{P_{\addressID_#1}}}
\newcommand\addressEdgePairs[2]{\ensuremath{P_{\addressID_#1,\addressID_#2}}}

\newcommand\entityID{\ensuremath{e}\xspace}
\newcommand\entityIDn[1]{\ensuremath{\entityID_#1}\xspace}
\newcommand\entityPairs[1]{\ensuremath{P_{\entityID_#1}}}
\newcommand\entityEdgePairs[2]{\ensuremath{P_{\entityID_#1,\entityID_#2}}}

\newcommand\aTransactionSet[2]{\ensuremath{T_{\addressID_#1,\addressID_#2}}\xspace}
\newcommand\eTransactionSet[2]{\ensuremath{T_{\entityID_#1,\entityID_#2}}\xspace}

\definecolor{AITturquoise}{RGB}{64,170,155}
\definecolor{AITbordeaux}{RGB}{121,11,26}
\definecolor{bluekeywords}{rgb}{0.13, 0.13, 1}
\definecolor{greencomments}{rgb}{0, 0.5, 0}
\definecolor{redstrings}{rgb}{0.9, 0, 0}
\definecolor{graynumbers}{rgb}{0.5, 0.5, 0.5}

\lstdefinestyle{mystyle}{
   	autogobble,
    columns=fullflexible,
    showspaces=false,
    showtabs=false,
    breaklines=true,
    showstringspaces=false,
    breakatwhitespace=true,
    escapeinside={(*@}{@*)},
    commentstyle=\color{greencomments},
    keywordstyle=\color{bluekeywords},
    stringstyle=\color{redstrings},
    numberstyle=\color{graynumbers},
    basicstyle=\ttfamily\footnotesize,
    frame=l,
    framesep=12pt,
    xleftmargin=12pt,
    tabsize=4,
    captionpos=b
}

\newcommand{\gsversion}{0.4.5\xspace}

\begin{document}
\title{GraphSense: A General-Purpose Cryptoasset Analytics Platform}

%
%
\author{
   Bernhard Haslhofer\inst{1} \and
   Rainer St\"{u}tz\inst{1} \and
   Matteo Romiti\inst{1} \and
   Ross King\inst{1}
}
\institute{
    AIT Austrian Institute of Technology\\
    Vienna, Austria
}

\maketitle              

\pagestyle{plain}

\noindent
\vspace{10pt}
\makebox[\linewidth]{Version \gsversion}


There is currently an increasing demand for cryptoasset analysis tools among cryptoasset service providers, the financial industry in general, as well as across academic fields.
At the moment, one can choose between commercial services or low-level open-source tools providing programmatic access.
In this paper, we present the design and implementation of another option: the GraphSense Cryptoasset Analytics Platform, which can be used for interactive investigations of monetary flows and, more importantly, for executing advanced analytics tasks using a standard data science tool stack.
By providing a growing set of open-source components, GraphSense could ultimately become an instrument for scientific investigations in academia and a possible response to emerging compliance and regulation challenges for businesses and organizations dealing with cryptoassets.



\section{Introduction}
\label{sec:introduction}

In recent years, we have observed a rapidly increasing demand for cryptoasset analysis tools in industry and academia: businesses dealing with cryptoassets analyze transactions to fulfill compliance guidelines and regulations (c.f.,~\cite{sackheim:2020,fatf:2019}); law enforcement needs these techniques to track and trace illicit money flows (e.g.,~\cite{PaquetClouson:2019aa,PaquetClouson:2019bb}); designers of distributed ledger technology analyze deployed systems to make informed system design decisions~\cite{Stuetz:2020a}; business analysts and investors analyze transactional data to understand markets; and, last but not least, scientists from a wide range of academic disciplines use cryptoasset analytics tools to find answers to their research questions.

At the moment, analysts can choose from two main options. On one hand, they can use commercial service offerings and analyze cryptoasset addresses and transactions via provided user interfaces and APIs. Neglecting the relatively high service costs, this has the advantage of a low entry barrier and availability of so-called attribution tags, which associate cryptoasset addresses with real-world actors such as exchanges. Alternatively, one can use free, open-source blockchain analytics tools like BlockSci~\cite{Kalodner:2020a}, which provides programmatic access to the full blockchain data and a highly efficient in-memory \emph{transaction graph} representation.

In this paper, we present a third option: the \emph{GraphSense Cryptoasset Analytics Platform}, which is designed as an extensible and scalable analytics platform for running customized analytics tasks on data gathered from multiple blockchains and other contextually relevant sources, such as exchange rate services. Similar to commercial offerings, GraphSense also provides a dashboard for basic, interactive investigations, which lowers the entry barrier for non-expert users. Similar to BlockSci, it provides the flexibility to perform analytics tasks on pre-computed graph abstractions. However, in contrast to BlockSci, GraphSense provides access to the so-called \emph{address and entity graphs}, which reflect the main structural elements of cryptoasset ecosystems: actors, who interact with each other and are linked together through cryptoasset transfers (c.f.,~\cite{Reid:2013aa}). Furthermore, GraphSense introduces the notion of \emph{TagPacks}, which support collaborative collection and provenance-aware curation of attribution tags, which are valuable data points in most analytics tasks.

Our vision was for GraphSense to become a general-purpose cryptoasset analytics platform that supports analysts in conducting microscopic, transaction-level investigations as well as more extensive macroscopic investigations on structural and dynamic aspects of cryptoasset ecosystems. Technically, GraphSense contributes reusable building blocks that can easily be integrated into an ETL or cryptoasset analytics pipeline. By being published\footnote{\url{https://github.graphsense.info}} under an open-source license, which permits reuse for commercial and non-commercial purposes, GraphSense has already attracted interest and contributions from third parties and could ultimately become a core technology for cryptoasset analytics research in academia and industry.

In the following, in Section~\ref{sec:background}, we first provide some background information on graph-abstractions required for cryptoasset analytics and then present our rationale for designing GraphSense. We then present the technical design and the architecture of the GraphSense platform in Section~\ref{sec:design}, before we provide further details on TagPacks in Section~\ref{sec:tagpacks}. Finally, in Section~\ref{sec:discussion}, we provide some insight into known challenges and future development directions.

This paper currently describes version \gsversion of the GraphSense Cryptoasset Analytics Platform. It will be updated based on users' feedback and new features included in future releases.


\section{Background and Design Rationale}
\label{sec:background}

\subsection{Cryptoasset Analytics}

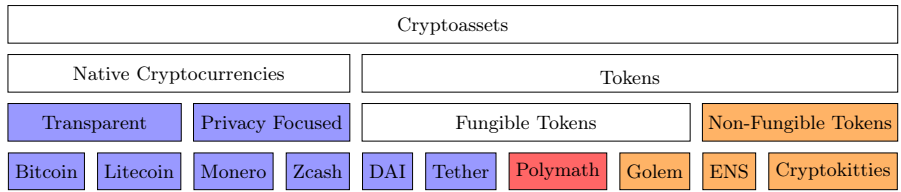
\begin{figure}
  \centering
  \adjustbox{max width=\columnwidth}{%
    \begin{tikzpicture}
	[extendrect/.style 2 args = {
		draw=black, text centered, 
		inner sep=0pt, outer sep=0pt, 
		fit=(#1) (#2)}]
	\tikzstyle{rect} = [rectangle, text centered, draw=black, minimum height = .6cm]
	
	\newcommand\Xdelta{.2cm}
	\newcommand\Ydelta{.8cm}

	\node [rect, fill=blue!40](zcash) at (0, 0) {Zcash};
	\node [rect, left = \Xdelta of zcash, fill=blue!40] (monero) {Monero};
	\node [rect, left = \Xdelta of monero, fill=blue!40] (litecoin) {Litecoin};
	\node [rect, left = \Xdelta of litecoin, fill=blue!40] (bitcoin) {Bitcoin};
	\node [rect, right = \Xdelta of zcash, fill=blue!40] (dai) {DAI};
	\node [rect, right = \Xdelta of dai, fill=blue!40] (tether) {Tether};
	\node [rect, right = \Xdelta of tether, fill=red!60] (polymath) {Polymath};
	\node [rect, right = \Xdelta of polymath, fill=orange!60] (golem) {Golem};
	\node [rect, right = \Xdelta of golem, fill=orange!60] (ens) {ENS};
	\node [rect, right = \Xdelta of ens, fill=orange!60] (cryptokitties) {Cryptokitties};

	\coordinate (privacyBL) at ($(monero.south west)+(0,\Ydelta)$);
	\coordinate (privacyTR) at ($(zcash.north east)+(0,\Ydelta)$);
	\node [extendrect={privacyBL}{privacyTR}, fill=blue!40, label={[shift={(0, -.6cm)}]Privacy Focused}] (privacy) {};

	\coordinate (transparentBL) at ($(bitcoin.south west)+(0,\Ydelta)$);
	\coordinate (transparentTR) at ($(litecoin.north east)+(0,\Ydelta)$);
	\node [extendrect={transparentBL}{transparentTR}, fill=blue!40, label={[shift={(0, -.6cm)}]Transparent}] (transparent) {};

	\coordinate (fungibleBL) at ($(dai.south west)+(0,\Ydelta)$);
	\coordinate (fungibleTR) at ($(golem.north east)+(0,\Ydelta)$);
	\node [extendrect={fungibleBL}{fungibleTR}, label={[shift={(0, -.6cm)}]Fungible Tokens}] (fungible) {};

	\coordinate (non-fungibleBL) at ($(ens.south west)+(0,\Ydelta)$);
	\coordinate (non-fungibleTR) at ($(cryptokitties.north east)+(0,\Ydelta)$);
	\node [extendrect={non-fungibleBL}{non-fungibleTR}, fill=orange!60, label={[shift={(0, -.6cm)}]Non-Fungible Tokens}] (non-fungible) {};

	\coordinate (nativeBL) at ($(transparent.south west)+(0,\Ydelta)$);
	\coordinate (nativeTR) at ($(privacy.north east)+(0,\Ydelta)$);
	\node [extendrect={nativeBL}{nativeTR}, label={[shift={(0, -.6cm)}]Native Cryptocurrencies}] (native) {};

	\coordinate (tokenBL) at ($(fungible.south west)+(0,\Ydelta)$);
	\coordinate (tokenTR) at ($(non-fungible.north east)+(0,\Ydelta)$);
	\node [extendrect={tokenBL}{tokenTR}, label={[shift={(0, -.6cm)}]Tokens}] (token) {};

	\coordinate (virtualBL) at ($(native.south west)+(0,\Ydelta)$);
	\coordinate (virtualTR) at ($(token.north east)+(0,\Ydelta)$);
	\node [extendrect={virtualBL}{virtualTR}, label={[shift={(0, -.6cm)}]Cryptoassets}] (virtual) {};

\end{tikzpicture}
  }
  \caption[]{
    The spectrum of cryptoassets. Technically, one can distinguish between \emph{Native Cryptocurrencies} and \emph{Tokens} deployed on platforms like Ethereum. From a usage perspective, one can distinguish between
    Payment Tokens
    (\tikz \fill[color=blue!40] (0,.6) rectangle (.7cm, .5cm);), 
    Security Tokens
    (\tikz \fill[color=red!60] (0,.6) rectangle (.7cm, .5cm);), and
    Utility Tokens
    (\tikz \fill[color=orange!60] (0,.6) rectangle (.7cm, .5cm);).
  }
  \label{fig:cryptoassets}
\end{figure}

In our terminology, we denote an \emph{asset} as something that has some value for someone. Building on this, we denote a \emph{cryptoasset} as a virtual asset that utilizes cryptography and some form of ledger technology, which may be distributed or not, for recording and sharing value transfers. As depicted in Figure~\ref{fig:cryptoassets}, we can roughly divide the spectrum of cryptoassets into \emph{native cryptocurrencies} like Bitcoin, and \emph{tokens} as they are deployed on account-model ledgers like Ethereum.

A \emph{cryptoasset ecosystem} represents a community of actors, who interact as a system and are linked together through cryptoasset transfers. 
The goal of \emph{cryptoasset analytics} is to develop and apply quantitative methods to understand the technical and socio-economic aspects of cryptoasset ecosystems. This has been our underlying motivation for building GraphSense.

The required algorithmic building blocks for enabling cryptoasset analytics depend on the conceptual design of a distributed ledger. Ledgers that follow Bitcoin's \emph{unspent transaction outputs} (UTXO) model, which includes Bitcoin derivatives like Litecoin and Zcash, allows a single transaction to have multiple inputs and outputs. We can compute various types of graph abstractions from the underlying blockchain~\cite{Reid:2013aa} such as the \emph{transaction graph}, which is a directed temporal graph connecting transactions by their inputs and outputs. Further, we can compute the \emph{address graph}, a bi-directed cyclic graph in which a node represents an address and an edge represents the set of transactions two addresses were involved in as input or output. Addresses can further be linked using various \emph{address clustering heuristics}, most importantly the so-called multiple-input or co-spent heuristic, which groups addresses that are likely controlled by the same real-world actor based on common use and reuse in transactions~\cite{Meiklejohn:2013aa}. After applying the clustering algorithm, one can build another graph abstraction: the so-called \emph{entity-graph}, which is a bi-directed, cyclic graph in which a node represents the set of addresses that are likely controlled by the same real-world actor (e.g., an exchange) and an edge that represents the aggregate set of transactions between two address sets (entities).

Other ledgers like Ethereum, NEO, or EOS follow a different conceptual model called the \emph{account model}. In that model, a single transaction has exactly one source and one destination account address. While it is still possible to compute the transaction and address graph, existing heuristics based on multiple inputs or outputs cannot be used. However, recent work has shown that address clustering is also possible for Ethereum's account model based on heuristics derived from the analysis of usage patterns surrounding deposit addresses, airdrops, or token transfer authorization~\cite{Victor:2020b}.

\subsection{Design Rationale}

The current system design of GraphSense reflects a number of observations and requirements that have registered over the past several years.

\paragraph{Data sovereignty} We observe that programmatic access to the full data, which includes blockchain data as well as attribution tags, is essential for efficient and effective analytics going beyond following individual transactions. When analyzing entire markets, as we did for Ransomware~\cite{PaquetClouson:2019aa} or Sextortion~\cite{PaquetClouson:2019bb}, one must extract relevant data points related to thousands of addresses in a single analytics task. Predefined interfaces, whether graphical dashboards or programmatic REST APIs, always somehow limit the scope of cryptocurrency analyses. Therefore, GraphSense follows a full data sovereignty strategy and provides programmatic access to the full underlying data, and thereby supports advanced usage scenarios.

\paragraph{Pre-computed Graph Abstractions} Nowadays, cryptoasset transactions can be inspected with a wide variety of publicly available blockchain explorers such as \texttt{blockchain.com}\footnote{\url{https://www.blockchain.com}} or \texttt{etherscan.io}\footnote{\url{https://etherscan.io}}. Next to providing details on individual blocks and transactions they also support navigation along the \emph{transaction graph}, which means users can navigate from a certain transaction output address to the next transaction that uses that address as input and vice versa. While this is certainly useful and important, we observed that many analytics tasks focus on the investigation of monetary flows between cryptoasset addresses, or more importantly between the real-world actors (e.g., exchanges) that somehow control these addresses, hence the cryptoasset entities. Therefore, GraphSense provides higher-level graph abstractions, namely the \emph{address- and entity-graph} for various cryptoassets. Since computing these graph abstractions is computationally expensive, GraphSense pre-computes them and makes them available for subsequent analytics tasks.

\paragraph{Collaborative Address Tagging} We are aware that \emph{attribution tags}, which associate cryptoasset addresses and entities with some real-world actor, are essential for conducting effective analyses. They can, for instance, be collected manually by interacting with certain services and assigning human-readable labels to addresses controlled by these services. Since this is a costly and resource-intensive task, we propose TagPacks, which is a simple file-structure for organizing and exchanging attribution tags. TagPacks can be collected and collaboratively maintained using Git\footnote{\url{https://git-scm.com}}, which is a free version control system that provides the technical means for recording data-provenance information. This aspect is important because documented evidence of the origin of data is increasingly emphasized or required in both academia and legal proceedings (c.f.~\cite{Froewis:2020a}).

\paragraph{Scalability \& Extensibility} We note that the volume of transactional data in blockchains is growing and new ledgers are appearing. In the long-run, in-memory graph representations might face \emph{scalability} issues when working with higher-level graph abstractions computed over several ledgers. We also follow the reasoning behind BlockSci and point out that most of the relevant data come from append-only data structures, which makes the ACID properties of general-purpose databases unnecessary. Therefore, we decided to build GraphSense on-top of a standard data science technology stack, which uses Apache Cassandra\footnote{\url{https://cassandra.apache.org}} as a NoSQL storage engine and Apache Spark\footnote{\url{https://spark.apache.org}} as an analytics engine. Both technologies can increase capacities by connecting additional hardware and therefore respond to growing data volumes. We also take into account that cryptoasset analysis is a highly dynamic field in which new ledgers are constantly being added or existing ones are changing. Therefore, we designed GraphSense as a modular and extensible analytics pipeline consisting of multiple, standalone building blocks, which can be updated, extended, and possibly replaced as needed.

\paragraph{Transparency \& Open Source} GraphSense leverages other open source efforts like BlockSci and uses them in an integrated analytics pipeline. In return, all GraphSense components are published as open-source software on GitHub under an MIT license\footnote{\url{https://opensource.org/licenses/MIT}} and can be re-used for commercial and non-commercial purposes. In this manner, GraphSense also fulfills the requirement of algorithmic transparency, which is another important condition for safeguarding the evidential value of cryptoasset investigations.

\section{Design and Architecture}\label{sec:design}

\subsection{Overall Architecture}

\begin{figure}
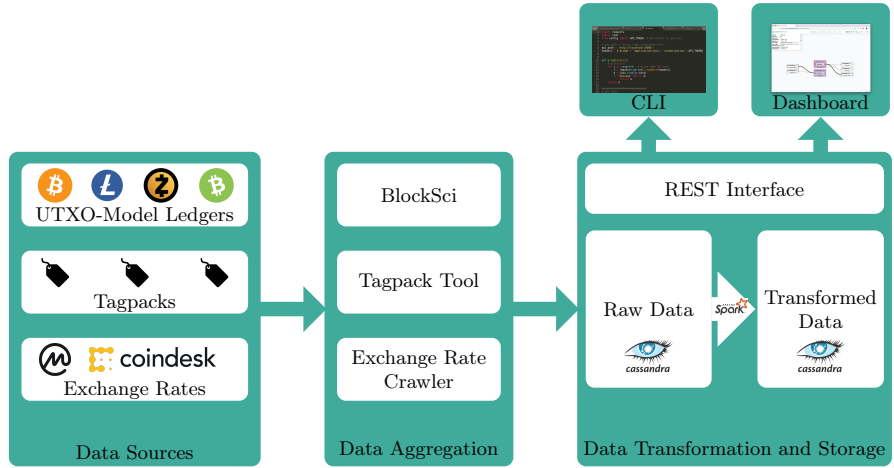

  \centering
  \adjustbox{max width=\columnwidth}{%
  \begingroup%
  \newcommand{\imgpath}{figures/graphsense-architecture}%
  \includestandalone[]{\imgpath/graphsense-architecture}%
  \endgroup

  }
  \caption{GraphSense Architecture}
  \label{fig:architecture}
\end{figure}

GraphSense is designed as a modular and extensible analytics pipeline consisting of multiple, standalone building blocks, which are connected and orchestrated via Docker\footnote{\url{https://www.docker.com}} and Docker Compose. As depicted in Figure~\ref{fig:architecture}, the overall pipeline can be divided into several parts: the relevant \emph{data sources} that provide the raw data points needed for further analyses; several \emph{data aggregation} components that retrieve data from different sources and ingest them into GraphSense's NoSQL storage back-end; a \emph{data transformation} job that computes statistical properties, clusters addresses and the required address- and entity graph abstractions; and finally interfaces that provide \emph{programmatic access} to the underlying data as well as a \emph{Dashboard} that supports users in analyzing individual nodes and edges in these graphs.

\subsection{Data Sources}

GraphSense uses data from the following sources:

\begin{itemize}
  
  \item \emph{UTXO-Model Ledgers}: At the moment, Bitcoin, Bitcoin Cash, Zcash, Litecoin are supported.

  \item \emph{TagPacks}: GraphSense integrates collaboratively collected attribution tags in the form of TagPacks. Further details will be provided in Sectionµ~\ref{sec:tagpacks}.

  \item \emph{Exchange Rates}: GraphSense utilizes cryptoasset exchange rates from public services such as CoinDesk\footnote{\url{https://www.coindesk.com}}, CoinMarketCap\footnote{\url{https://coinmarketcap.com}}, and the European Central Bank (ECB)

\end{itemize}

\subsection{Data Aggregation}

Raw data are aggregated from the above sources using several
data-source-specific connectors and extractors. Table~\ref{tab:data-summary} shows the number of blocks, transactions, addresses, and tags that are aggregated at the time of this writing.

\begin{table*}[h]
  \centering%
  \caption{Summary of supported cryptocurrency ledgers.}%
  \begin{tabular*}{\textwidth}{l@{\extracolsep{\fill}}r@{\extracolsep{\fill}}r@{\extracolsep{\fill}}r@{\extracolsep{\fill}}r@{\extracolsep{\fill}}r}
  \toprule
Currency & Date & \#Blocks & \#Transactions & \#Addresses & \#Tags \\ 
  \midrule
BTC & 2021-02-19 & 671,193 & 617,563,419 & 786,122,678 & 5,858 \\ 
  BCH & 2021-02-19 & 675,552 & 304,347,192 & 315,176,329 & 32 \\ 
  LTC & 2021-02-19 & 2,003,864 & 61,025,669 & 66,969,867 & 42 \\ 
  ZEC & 2021-02-19 & 1,153,493 & 7,990,718 & 5,270,129 & 15 \\ 
   \bottomrule
\end{tabular*}
  \label{tab:data-summary}%
\end{table*}

For UTXO model ledgers, GraphSense currently relies on BlockSci, which provides an efficient parser for large chains like Bitcoin as well as REST-API connectors for other, smaller ledgers. At the moment, GraphSense also relies on BlockSci's mapping from transaction and address hashes to integer IDs, which significantly lowers memory consumption and storage space.

Bitcoin exchange rates are gathered from CoinDesk's public API (Bitcoin Price Index API\footnote{\url{https://www.coindesk.com/API}}), where historical exchange rates are provided for different FIAT currencies in JSON format at a dedicated API endpoint\footnote{\url{https://api.coindesk.com/v1/bpi/historical/close.json}}. Daily historical exchange rates for the remaining cryptocurrencies are retrieved in U.S.\ Dollars from the CoinMarketCap API\footnote{\url{https://web-api.coinmarketcap.com/v1/cryptocurrency/ohlcv/}}. For conversion to other fiat currencies we use foreign exchange rates provided by the European Central Bank (ECB)\footnote{\url{https://www.ecb.europa.eu/stats/eurofxref/eurofxref-hist.zip}}.

TagPacks can be aggregated and validated using the GraphSense TagPack Management Tool. Since TagPacks use terms from external taxonomies\footnote{\url{https://interpol-innovation-centre.github.io/DW-VA-Taxonomy}}, that tool can also be used for aggregating, validating and ingesting taxonomy concepts and definitions.

All aggregated raw data are ingested into the NoSQL storage back-end in a dedicated \emph{raw keyspace}.

\subsection{Transformation}

The next step in the GraphSense data analytics pipeline is a transformation job that computes statistical summaries on central blockchain entities (blocks, transactions, addresses). For UTXO ledgers, it also computes clusters of addresses that are likely controlled by the same real-world entity, which could, for instance, be an exchange. The entire transformation job is implemented in Apache Spark and runs in parallel over raw data items, which are stored in Apache Cassandra and distributed over a cluster of connected machines.

\paragraph{Statistical properties} The properties computed for blocks and transactions are trivial and roughly correspond to those that can also be found in public blockchain explorers (e.g., total transaction inputs and outputs). For addresses and entities, however, we compute semantically richer statistics such as the total volume of currency units received by an address, while taking into account historical exchange rates for each transaction.

\paragraph{Address graph} A cryptoasset address \addressIDn{i} represents a node in the address graph and carries a set of key-value pairs \addressPairs{i} providing statistical summaries for individual addresses: number of i) deposits, ii) withdrawals, iii) depositing addresses, iv) withdrawing addresses, v) coins received, vi) coins spent and vii) balance as well as viii) activity period based on the ix) first transaction and the x) last transaction.

The aggregated set of transactions \aTransactionSet{i}{j} from address \addressIDn{i} to \addressIDn{j} represents the edge between the two nodes and is also labeled with key-value pairs \addressEdgePairs{i}{j}: i) estimated transferred value, ii) number of transactions and iii) list of transactions. Here we point out that an exact computation of the value transfer between two addresses in UTXO ledgers is not possible, because a single UTXO transaction has multiple inputs and outputs. Therefore, it is not possible to associate a value from one input address with an output address (see~\cite{Haslhofer:2016ab}).

Since each address node carries several computed statistical properties and also the edges are labeled with properties and values, we represent the address graph following the property graph model~\cite{Rodriguez:2010}. A property graph is essentially a bi-directed multi-graph with labeled nodes and edges, where edges have their own identity.

\paragraph{Address clustering} In UTXO ledgers, a user can create and control an arbitrary number of addresses at virtually no cost. Linking and clustering these addresses into a single set, which represents the real-world entity that likely controls these addresses, is an essential task in cryptoasset analytics. GraphSense currently implements the co-spent heuristics~\cite{Meiklejohn:2013aa}, which is also known as multiple-input heuristics and assumes that inputs spent in the same transactions are controlled by the same user who must possess the corresponding private key for signing these inputs. While this method has proved very effective in practice~\cite{Harrigan:2016a}, a known, possible source for false positives are CoinJoins, which can be identified and filtered before applying that heuristics (see~\cite{Kalodner:2020a}). Other clustering heuristics rely on the identification of change addresses in the transaction outputs. Since this depends on the technical nature of the client executing the transactions, GraphSense refrains from implementing any change heuristics.

From a technical perspective, clustering is therefore implemented as a union-find algorithm that selects all address IDs from all non-multi-signature transactions with more than one input, ships them to a central master node where the disjoint-set data structures are computed, and ships them back to all nodes in the cluster for assigning unique cluster or entity IDs to each address.

\paragraph{Entity graph} By combining the previously described address graph with the entities (disjoint address sets) computed by address clustering, we can now build the \emph{entity graph}. In the entity graph, a node represents an entity \entityIDn{x} which reflects some real-world actor (e.g., an exchange) controlling a set of addresses, while an edge represents the aggregated set of transactions \eTransactionSet{x}{y} that occurred between two entities \entityIDn{x}, \entityIDn{y}.

In general, the entity graph carries the same properties as the address graph, but on an aggregated level. Hence, a node \entityIDn{x} carries the following key-value pairs \entityPairs{x}: number of i) deposits, ii) withdrawals, iii) depositing entities, iv) withdrawing entities, v) coins received, vi) coins spent, vii) balance as well as viii) activity period based on the ix) first transaction and the x) last transaction and, additionally, xi) the number of addresses and xii) a tag coherence score\footnote{Tag coherence: a metric that uses the string similarity between tags related to the entity addresses to describe the entity consistency and composition.}.

Analogously, an edge has the following aggregated key-value pairs \entityEdgePairs{x}{y}: i) estimated transferred value, ii) number of transactions and iii) list of transactions\footnote{Since storing the entire list of transactions among two entities might be expensive, we disregard transaction lists with more than 100 entries.}.

\begin{figure}
  \centering
  \begin{tikzpicture}
	\newcommand\ARadio{1cm}
	\newcommand\ERadio{1.95cm}

	\newcommand\Ydistance{2cm}
	\newcommand\Xdistance{4cm}
	\newcommand\LineWidth{.3mm}
	\newcommand\XLabelShift{.65cm}
	\newcommand\YLabelShift{-.45cm}
	\newcommand\eYPropShift{-.175cm}
	\newcommand\eXPropShift{1.3cm}
	\newcommand\aYPropShift{-.2cm}
	\newcommand\aXPropShift{.55cm}

	\tikzstyle{anode} = [circle, text centered, draw=black, minimum size = \ARadio, fill=AITbordeaux!40]
	\tikzstyle{enode} = [rectangle, rounded corners, text centered, draw=black, minimum height = \ERadio, minimum width = \ERadio, fill=AITturquoise!40]

	\node [anode](a1) at (0, 0) {\addressIDn{1}};
	\node [anode, below = of a1](a2) {\addressIDn{2}};
	\node [anode, right = \Xdistance of a1](a3) {\addressIDn{3}};
	\node [anode, right = \Xdistance of a2](a4) {\addressIDn{4}};

	\begin{scope}[on background layer]
		\node [enode, minimum height = 2*\ERadio, label={[shift={(\XLabelShift, \YLabelShift)}]\entityIDn{1}}](e1) at ($(a1)!0.5!(a2)$) {};
		\node [enode, label={[shift={(\XLabelShift, \YLabelShift)}]\entityIDn{2}}](e2) at (a3) {};
		\node [enode, label={[shift={(\XLabelShift, \YLabelShift)}]\entityIDn{3}}](e3) at (a4) {};		
	\end{scope}
	\node [above = \eYPropShift of e1, label={[shift={(\eXPropShift, \eYPropShift)}]\{\entityPairs{1}\}}] (e1prop) {};
	\node [above = \eYPropShift of e2, label={[shift={(\eXPropShift, \eYPropShift)}]\{\entityPairs{2}\}}] (e2prop) {};
	\node [above = \eYPropShift of e3, label={[shift={(\eXPropShift, \eYPropShift)}]\{\entityPairs{3}\}}] (e3prop) {};

	\node [above = \eYPropShift of a1, label={[shift={(-\aXPropShift, \aYPropShift)}]\{\addressPairs{1}\}}] (a1prop) {};
	\node [above = \eYPropShift of a2, label={[shift={(-\aXPropShift, \aYPropShift)}]\{\addressPairs{2}\}}] (a2prop) {};
	\node [above = \eYPropShift of a3, label={[shift={(-\aXPropShift, \aYPropShift)}]\{\addressPairs{3}\}}] (a3prop) {};
	\node [above = \eYPropShift of a4, label={[shift={(-\aXPropShift, \aYPropShift)}]\{\addressPairs{4}\}}] (a4prop) {};

	\draw [color=AITbordeaux, text=black, line width=\LineWidth, ->] (a1) -- (a3) node [midway, above] {\{\addressEdgePairs{1}{3}\}};
	\draw [color=AITbordeaux, text=black, line width=\LineWidth, ->] (a2) -- (a4) node [midway, below] {\{\addressEdgePairs{2}{4}\}};
	\draw [color=AITbordeaux, text=black, line width=\LineWidth, ->] (a1) -- (a4) node [pos=.7, above, sloped] {\{\addressEdgePairs{1}{4}\}};
	\draw [color=AITturquoise, text=black, line width=\LineWidth, ->] (e1.east) -- (e2) node [midway, above, sloped] {\{\entityEdgePairs{1}{2}\}};
	\draw [color=AITturquoise, text=black, line width=\LineWidth, ->] (e1.east) -- (e3) node [pos=.3, below, sloped] {\{\entityEdgePairs{1}{3}\}};

\end{tikzpicture}
  \caption[]{
  Conceptual Address
  (\tikz \fill[color=AITbordeaux!40] (0,.6) rectangle (.7cm, .5cm);)
  and Entity
  (\tikz \fill[color=AITturquoise!40] (0,.6) rectangle (.7cm, .5cm);)
  Graph Model. The Entity Graph is a higher-level abstraction and an aggregated model of the Address Graph, both at node and edge level. In this example, \entityEdgePairs{1}{3} is an aggregation of \addressEdgePairs{1}{4} and \addressEdgePairs{2}{4}.
  }
  \label{fig:address-entity-graph}
\end{figure}
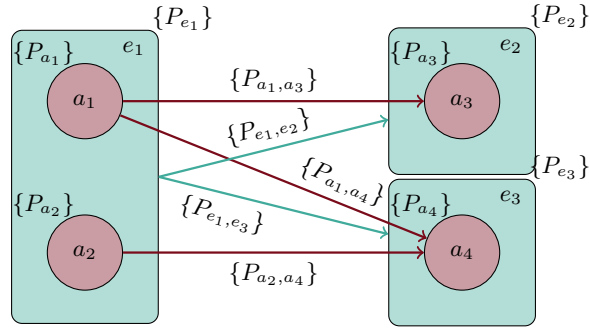

Figure~\ref{fig:address-entity-graph} illustrates both the address and entity property graphs. Addresses \addressIDn{1} and \addressIDn{2} are clustered into entity \entityIDn{1}, while entity \entityIDn{2} and \entityIDn{3} are made of one address only (\addressIDn{3} and \addressIDn{4}, respectively). Table~\ref{tab:graph-summary} shows the dimensionality (number of nodes and edges) of both graphs and one can clearly see that the entity graph reduces the dimensionality. In the entity graph, the number of nodes is approximately halved, and the number of edges is reduced by factor of 3.5--5, respectively.

\begin{table*}
  \centering%
  \caption{Summary of computed graph representations.}%
  \begin{tabular*}{\textwidth}{l@{\extracolsep{\fill}}r@{\extracolsep{\fill}}r@{\extracolsep{\fill}}r@{\extracolsep{\fill}}r}
  \toprule
 & \multicolumn{2}{c}{Address Graph} & \multicolumn{2}{c}{Entity Graph}\\
 \cline{2-3} \cline{4-5}
Currency & \#Nodes & \#Edges & \#Nodes & \#Edges \\ 
  \midrule
BTC & 786,122,678 & 4,802,966,573 & 372,206,870 & 942,849,825 \\ 
  BCH & 315,176,329 & 1,841,308,966 & 145,420,818 & 387,230,156 \\ 
  LTC & 66,969,867 & 310,001,795 & 32,234,112 & 88,659,704 \\ 
  ZEC & 5,270,129 & 67,480,350 & 2,552,050 & 13,847,404 \\ 
   \bottomrule
\end{tabular*}
  \label{tab:graph-summary}%
\end{table*}

\paragraph{Graph Storage}

In GraphSense, the address and entity graphs are stored as node and edge lists in a distributed NoSQL database. Since NoSQL databases typically don't support efficient lookup-by-indices on non-partition keys, GraphSense stores each edge list twice: once to support retrieval of an edge by source node id and once to support the reverse direction. While we consider the additional required disk space as being a non-issue, the challenge clearly lies in partitioning the edge list across machines so that partition sizes follow a roughly uniform distribution and the data keeps load balanced throughout the cluster.

\subsection{Programmatic Access}

A wide range of analyses, which go beyond an inspection of individual transactions, can only be achieved through programmatic access to the entire underlying dataset. Although this is associated with an additional effort initially, one gains reproducibility and repetition with minimal additional costs. Therefore, GraphSense offers two options for programmatic access: a REST-API and the possibility to run customized Apache Spark Jobs over the entire dataset.

The REST-API follows the OpenAPI specification\footnote{\url{https://swagger.io/specification}}, which defines a standard, language-agnostic interface to RESTful APIs that can be used by code generation tools to generate servers and clients in various programming languages. GraphSense implements a REST-Server-Stub in Python Flask and currently provides client libraries in Python and R.

The second more powerful option is to implement a customized Apache Spark job and run it over the entire dataset. This of course requires direct access to the cluster running GraphSense and full knowledge and understanding of the NoSQL model, which is used for storing the data.

\subsection{Dashboard}

\begin{figure}
  \adjustbox{max width=\textwidth}{%
    \includegraphics{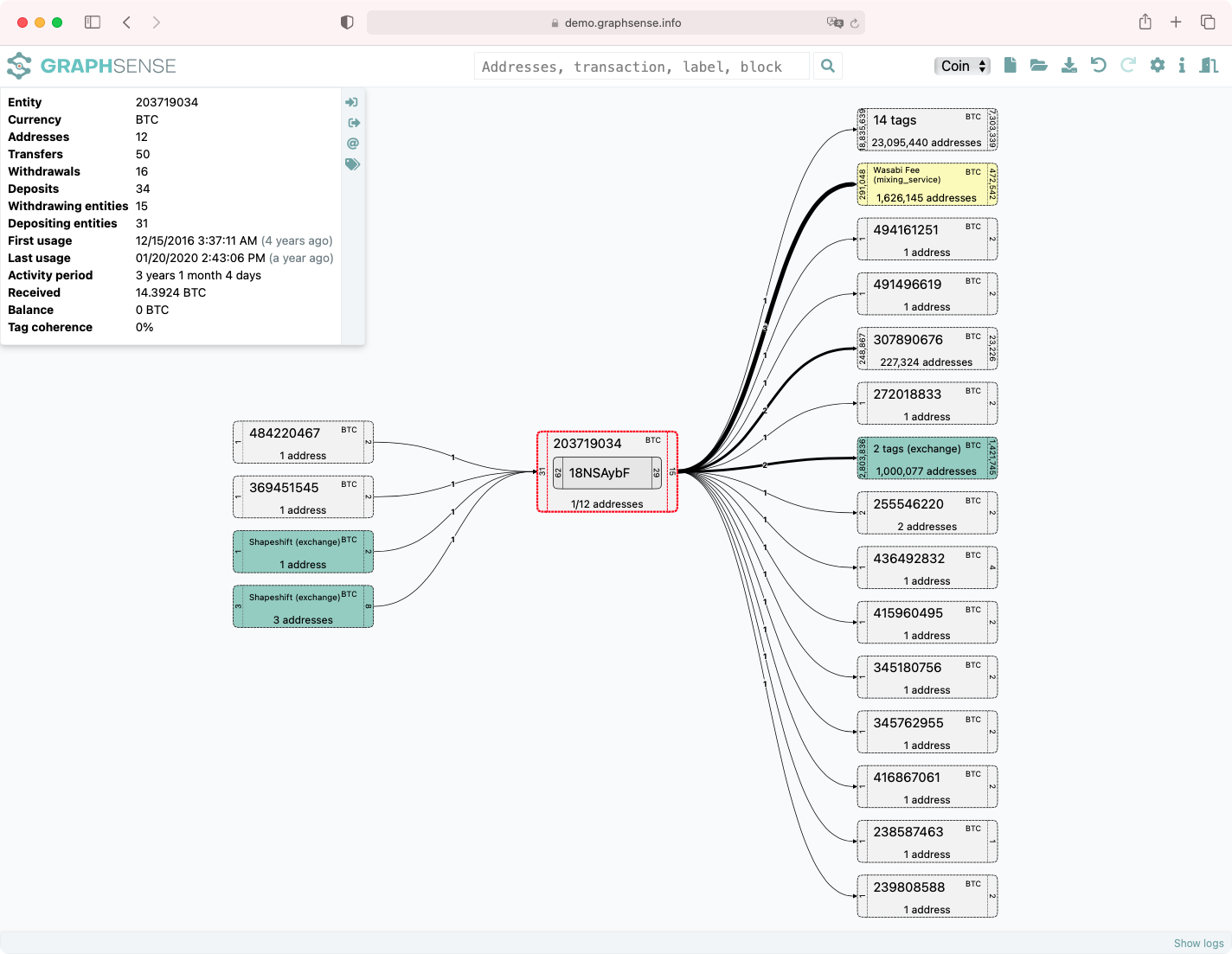}
  }
  \caption{Screenshot of the GraphSense Dashboard}
  \label{fig:graphsense_dashboard}
\end{figure}

In order to provide a low entry barrier for non-expert users, GraphSense also provides a visual Dashboard, as shown in Figure~\ref{fig:graphsense_dashboard}. It supports the inspection of blocks, transactions, addresses, and entities as well as navigation along the nodes and edges of the address and entity graph. In this manner, users can trace monetary flows and construct relevant sub-graphs reflecting the result of their investigations. The dashboard also provides means for automatically searching for certain types of nodes, such as entities representing exchanges, within certain boundary conditions (e.g., maximum node degrees). Users can also annotate nodes, export graphs, import additional tags, and download audit logs of their interactions.

Technically, the Dashboard is implemented as a pure JavaScript REST-API client, which is bundled using webpack\footnote{\url{https://webpack.js.org}}. It is also important to emphasize that the GraphSense Dashboard is read-only, which means that no user interactions or data entered by the user are sent back to the GraphSense server.


\section{TagPacks}
\label{sec:tagpacks}

Attribution tags are any form of context information that can be attributed to an address, transaction, or cluster, such as the name of an exchange hosting the associated wallet or some other personally identifiable information (PII) of the account holder. The strength of the attribution approach lies in combining address clusters with attribution tags: a tag attributed to a single address being controlled by some cryptoasset service, which typically forms a large address cluster, can easily de-anonymize hundreds of thousands of addresses.

In our previous work~\cite{Froewis:2020a}, we have already highlighted the important role of attribution tags in modern cryptoasset analytics and identified key legal requirements for the forensic processing of data. We pointed out that the \emph{provenance} of attribution tags is a critical foundation for assessing their quality and authenticity, as well as for enabling trust and allowing reproducibility. If used for law enforcement purposes, the provenance even becomes a legal requirement.

In the following, we describe how we collect and organize attribution tags in so-called TagPacks, how we shared and collaboratively managed them using Git, and how we intend to establish attribution tag interoperability among tools by using an agreed-upon taxonomy.

\subsection{TagPack Structure}

A TagPack defines a structure for collecting and packaging attribution tags with additional shared provenance metadata (e.g., title, creator, etc.). TagPacks are represented as YAML files, which can easily be created by hand or exported automatically from other systems. Listing~\ref{lst:sample_tagpack} shows a minimal TagPack, which attributes addresses from different ledgers (BTC, BCH, ZEC) to the ``Internet Archive'', which is a non-profit organization in the US. It also records the \emph{creator} of this TagPack, its last modification date (\emph{lastmod}), the type of entity controlling these addresses (\emph{category}), as well as the information \emph{source}.

\lstset{style=mystyle}
\begin{lstlisting}[
	label={lst:sample_tagpack},
	caption={Minimal TagPack Example}]
title: GraphSense Demo TagPack
creator: GraphSense Team
description: A collection of tags commonly used for demonstrating GraphSense features
label: Internet Archive
category: organization
lastmod: 2019-06-12
source: https://archive.org/donate/cryptocurrency
tags:
- address: 1Archive1n2C579dMsAu3iC6tWzuQJz8dN
  currency: BTC
- address: 1K1rgZ1dz9w7dsR1HGS1drmzfUHMtqx1Tc
  currency: BCH
- address: t1ZmpK4QFcvyQZ3ghTgSboBW8b4HgiZHQF9
  currency: ZEC
\end{lstlisting}

A TagPack consists of a \emph{header} and a \emph{body} section. The header lists several mandatory and optional metadata fields and the body provides the list of tags. The range of possible properties for header and body entries is defined in a TagPack Schema. In the above example, the properties \emph{title} and \emph{creator} are part of the TagPack header, the list of tags represents the body.

To avoid that property values need to be repeated for all tags in a TagPack, body fields can also be abstracted and added to the header, thereby being inherited by all body elements, as shown in the following example. In the example above, the label \emph{Internet Archive} is a body-level tag property, which has been added to the header to avoid repetition. It is also possible to override abstracted fields in the body. This could be relevant if someone creates a TagPack comprising several tags and then adds additional tags later on, which then, of course, have different property values.

GraphSense also provides a dedicated TagPack Management Tool\footnote{\url{https://github.com/graphsense/graphsense-tagpack-tool}}, which allows validation of TagPacks against the TagPack schema and referenced taxonomies before they are ingested into the NoSQL storage back-end and processes as part of the transformation step.

\subsection{Collaborative Tag Sharing}

Instead of defining and building a data provenance model and management system from scratch, GraphSense adopts Git for storing and publishing attribution tags. Git has its origin in distributed software development and has, over the last decade, become the de-facto standard for publishing and tracking changes in source code files. It automatically creates hashes over each file and allows, if required, users to digitally sign their contents after each commit. Git is increasingly used for sharing smaller and even large datasets (Git LFS\footnote{\url{https://git-lfs.github.com}}).

\subsection{Attribution Tag Interoperability}

The use of common terminologies is essential for data sharing and establishing interoperability across tools. Therefore, the TagPack schema defines two properties that take concepts from agreed-upon taxonomies as values:

\begin{itemize}

	\item \emph{category}: defines the type of real-world entity that is in control of a given address. Possible concepts (e.g., Exchange, Marketplace) are defined in the INTERPOL Darkweb and Cryptoassets Entity Taxonomy\footnote{\url{https://github.com/INTERPOL-Innovation-Centre/DW-VA-Taxonomy\#entities}}.

	\item \emph{abuse}: if an address was involved in some abusive behavior, this property's value defines the type of abuse and can take values from the INTERPOL Darkweb and Cryptoassets Abuse Taxonomy\footnote{\url{https://github.com/INTERPOL-Innovation-Centre/DW-VA-Taxonomy\#abuse}}.

\end{itemize}

In the example TagPack provided in Listing~\ref{lst:sample_tagpack}, for instance, the real world actor controlling these addresses is categorized as \emph{organization}, which directly maps to a concept that is uniquely identified via a URI\footnote{\url{https://interpol-innovation-centre.github.io/DW-VA-Taxonomy/taxonomies/entities\#organization}} as part of the INTERPOL Darknet Entity Taxonomy. If all cryptoasset analytics tools categorize attribution tags according to these taxonomies and also use the provided definitions, then attribution data can be harmonized across tools and the first step towards better interoperability can be achieved.

\section{Discussion}\label{sec:discussion}

Given current developments in the cryptoasset field, we strongly believe that there will a need for a deeper quantitative understanding of both individual and aggregate transaction flows, and of the technical and socio-economic aspects of increasingly complex cryptocurrency ecosystems. Networks are natural abstractions for such systems as they provide the basis for task-specific measurement and simulation methods. With GraphSense we provide the required computational infrastructure and pre-computed network abstractions for implementing such methods. With its modular, horizontally scalable system architecture, GraphSense also provides the flexibility to quickly react to upcoming, yet unforeseen developments and methodological challenges in this field.

\paragraph{Limitations} GraphSense is a steadily evolving system and also faces some yet unresolved limitations. First, the price of horizontal scalability is that GraphSense runs on a distributed hardware infrastructure. The operation of such an infrastructure requires a specific skill-set, which is hard to find and also requires relatively large initial investment costs. However, we argue that hosting GraphSense externally (e.g., in commercial cloud infrastructure) might become even more costly with increasing data volumes and involve yet unforeseen technical, organizational, and financial dependencies.

Lack of real-time updates is another inherent limitation of the overall system architecture, which has been designed for data analytics workflows. The bottleneck lies in updating the address clusters and in re-computing the graph abstractions, which can, depending on the dimensions of the hardware cluster, take several hours. However, we argue that real-time investigations are hardly ever needed, because most analytics tasks, for instance, the forensic analysis of a ransomware attack, are conducted in retrospect.

The third limitation we are facing at the moment is the lack of incentive for collecting and sharing attribution tags. The industry has the means for collecting but not the incentive for sharing for competitive reasons; in academia, there is an incentive for sharing, for scientific reproducibility for example, but typically few resources for collecting.

Finally, we also would like to point out that the overall analytics pipeline could still be optimized. Address clusters, for instance, are currently computed centrally, which involves unnecessary communication costs. Alternatively, one could use partition-aware connected component detection algorithms, which promise to be more efficient~\cite{park:2020a} but have not yet been evaluated within GraphSense.

\paragraph{Outlook} GraphSense follows an agile release plan with major and minor releases. With the next upcoming minor release (0.4.6), it should be possible to deploy GraphSense on a single server and retrieve pre-computed dumps from a periodically updated data repository. The next major release (0.5.X) will support account model ledgers, starting with Ethereum. Depending on the adoption of off-chain payment channels, a future major release (0.6.X) might also support analysis of payment channel transactions across ledgers (c.f.~\cite{Romiti:2021a}).

We also envision GraphSense to become a key technology in a research sub-field, which we call \emph{CryptoFinance}\footnote{\url{https://www.csh.ac.at/complexity-science/Cryptofinance/}}. The goal is to systemically assess emerging technologies and paradigms like Decentralized Finance (DeFi), to learn more about opportunities and risks associated with these developments, and to ultimately come up with measures that help us in quantifying systemic risks in cryptoasset ecosystems. Efficient and effective computations over graph and network abstractions will certainly play a central role in this effort.


\section{Conclusions}
\label{sec:conclusions}

GraphSense is a response to the increasing need for a general-purpose cryptoasset analytics platform. In this paper, we discussed the network abstractions relevant in this field, elaborated on our design rationale, and described the architecture and current technical building blocks of GraphSense.
We will certainly continue developing GraphSense as part of our research activities in the field of cryptoasset analytics and expect that GraphSense will soon support analytics across UTXO and account-model ledgers and, if relevant, also off-chain payment channels. We also expect and see to some extent that future GraphSense development will become more a community effort, which is driven by academia and stakeholders in the financial industry and the emerging CryptoFinance field.

\bibliographystyle{splncs04}
\bibliography{references}

\end{document}